\documentclass{article}

\usepackage{PRIMEarxiv}
\usepackage[utf8]{inputenc} % allow utf-8 input
\usepackage[T1]{fontenc}    % use 8-bit T1 fonts
\usepackage{hyperref}       % hyperlinks
\usepackage{url}            % simple URL typesetting
\usepackage{booktabs}       % professional-quality tables
\usepackage{amsfonts}       % blackboard math symbols
\usepackage{nicefrac}       % compact symbols for 1/2, etc.
\usepackage{microtype}      % microtypography
\usepackage{lipsum}
\usepackage{fancyhdr}       % header
\usepackage{graphicx}       % graphics
\graphicspath{{media/}}     % organize your images and other figures under media/ folder

\title{Deep Learning Framework with Multi-Head Dilated Encoders for Enhanced Segmentation of Cervical Cancer on Multiparametric Magnetic Resonance Imaging

}

\author{
  Reza Kalantar\textsuperscript{1,2}, Sebastian Curcean\textsuperscript{2,3}, Jessica M Winfield\textsuperscript{1,2}, Gigin Lin\textsuperscript{4}, Christina Messiou\textsuperscript{1,2}, Matthew D Blackledge\textsuperscript{1,2,*} and Dow-Mu Koh\textsuperscript{1,2}
}

\begin{document}
\maketitle
\begin{abstract}
T\textsubscript{2}-weighted magnetic resonance imaging (MRI) and diffusion-weighted imaging (DWI) are essential components for cervical cancer diagnosis. However, combining these channels for training deep learning models are challenging due to misalignment of images. Here, we propose a novel multi-head framework that uses dilated convolutions and shared residual connections for separate encoding of multiparametric MRI images. We employ a residual U-Net model as a baseline, and perform a series of architectural experiments to evaluate the tumor segmentation performance based on multiparametric input channels and feature encoding configurations. All experiments were performed using a cohort including 207 patients with locally advanced cervical cancer. Our proposed multi-head model using separate dilated encoding for T\textsubscript{2}W MRI, and combined b1000 DWI and apparent diffusion coefficient (ADC) images achieved the best median Dice coefficient similarity (DSC) score, 0.823 (confidence interval (CI), 0.595-0.797), outperforming the conventional multi-channel model, DSC 0.788 (95\% CI, 0.568-0.776), although the difference was not statistically significant (p>0.05). We investigated channel sensitivity using 3D GRAD-CAM and channel dropout, and highlighted the critical importance of T\textsubscript{2}W and ADC channels for accurate tumor segmentations. However, our results showed that b1000 DWI had a minor impact on overall segmentation performance. We demonstrated that the use of separate dilated feature extractors and independent contextual learning improved the model's ability to reduce the boundary effects and distortion of DWI, leading to improved segmentation performance. Our findings can have significant implications for the development of robust and generalizable models that can extend to other multi-modal segmentation applications.
\end{abstract}

% keywords can be removed

\vspace{2mm} %2mm vertical space

%% ---- Introduction ----
\section{Introduction}
\setstretch{1.35} 
Multiparametric magnetic resonance imaging (mpMRI) is a crucial tool in the diagnosis and management of gynecological malignancies, including cervical cancer. It provides detailed anatomical and functional information, which is applied for disease staging, treatment planning, response monitoring, and surveillance for disease recurrence\cite{togashi1989carcinoma, green2001survival, pollard2017future}. An important aspect of many mpMRI protocols is diffusion-weighted imaging (DWI), which enhances the contrast and visualization of cellular tumours. DWI is sensitive to the rate of diffusion of water molecules \emph{in vivo}, which is hindered due to increased cell density \cite{koh2010diffusion}; the rate of diffusion can be quantified at each spatial location through estimation of maps of apparent diffusion coefficient (ADC). ADC measurements are being increasingly used as a surrogate biomarker of tumour grade \cite{yoshida2017dwi, tsuruta2022dwi}, and have shown promising results in identifying early treatment responses, making them desirable for monitoring therapeutic outcomes in cervical cancer \cite{higaki2018introduction, abd2020impact}. On conventional anatomical T\textsubscript{2}-weighted (T\textsubscript{2}W) MRI, primary and metastatic tumors exhibit intermediate to high signal intensities, which is used to identify cervical abnormalities, as well as for disease staging and directing MRI-guided interventions \cite{subak1995cervical, romesser2021magnetic}. 

Automating disease detection and delineation on medical images is a critical task, primarily because it aids in extracting valuable biomarkers from images, which enhances clinical decision-making. This process is currently challenged by the requirement for extensive annotated datasets, leading to a high dependency on clinicians and potential inconsistencies due to human contouring variations. Automated tumor segmentation on mpMRI, therefore, holds great significance, not only for reducing the burden on clinicians but also for its potential in improving accuracy and consistency. Furthermore, in contexts such as tumor planning, where manual delineation may not be feasible due to time constraints, these automated methods become particularly indispensable. Therefore, the development of fully automatic tumor segmentation techniques is a crucial step towards achieving more efficient and reliable clinical processes.

The advent of advanced imaging and high-performance technologies has led to a surge of interest in deep learning (DL) and convolutional neural network (CNN)-based approaches for pelvic cancer segmentation \cite{kalantar2021automatic}. While several studies have explored DL-based segmentation of cervical cancer on MRI \cite{rigaud2021automatic, ma2022deep, wang2021multimodal, lu2022augms, zabihollahy2022fully}, there remains a scarcity of research on cancer tumor segmentation utilizing multiparametric MRI. Specifically, few studies have combined semantic knowledge between T\textsubscript{2}W MRI and DWI/ADC in this context.

Among related studies, Lin et al. \cite{lin2020deep} developed a U-Net model for segmenting cervical cancer on DWI and found that multi-channel input (b0, b1000, and ADC) produced the best segmentation performance. However, this study focused on two-dimensional (2D) images and did not incorporate multimodal MRI (e.g. MRI images derived from different sequences). Kano et al. \cite{kano2021automatic} combined 2D and three-dimensional (3D) U-Net models for cervical tumor segmentation on DWI using an ensembling approach. Yoganathan et al. \cite{yoganathan2022automatic} segmented primary tumors along with organs-at-risk (OARs) on T\textsubscript{2}W MRI, reporting that integrating segmentations from 2.5D training in axial, coronal, and sagittal planes improved segmentation performance compared with previous 2D models. However, this study was limited to 39 patients and single-channel inputs. Wang et al. \cite{wang2021multimodal} proposed a 3D CNN model for cervical tumor segmentation on multimodal MRI, while Hodneland et al. \cite{hodneland2022fully} utilized a U-Net with residual connections, employing T\textsubscript{2}W MRI, b1000 DWI, and ADC as input channels. However, none of these studies examined the impact of spatial mismatch between multimodal MRI inputs, resulting from distortion in echo-planar imaging (EPI) and soft-tissue deformations between scans \cite{yoshizako2021comparison, donato2014geometric}, on cervical tumor segmentation outcome.

The aim of this study was to develop a novel 3D DL framework that included multi-head dilated residual encoding blocks for combined fine-grained and contextual feature aggregation and training on anisotropic sub-volumes of images, enhancing the segmentation of locally advanced cervical tumors on multiparametric MRI. To our best knowledge, no previous studies have investigated this strategy for automated segmentation of pelvic malignancies.

\vspace{5mm} %5mm vertical space

%% ------ Materials and Methods ------
\section{Materials and Methods}
\label{sec:headings}
\subsection{Patient Populations and Imaging Parameters}
For this study, a retrospective cohort consisting of 207 patients diagnosed with locally advanced cervical cancer and who underwent pelvic T\textsubscript{2}W MRI and DWI on a 3T MAGNETOM TrioTim MRI scanner (Siemens Healthcare, Erlangen, Germany) were selected. The ground truth tumor contours were defined by a clinician with 3 years of experience on T\textsubscript{2}W MRI images with the DWI data available for all patients. The acquisition parameters for this dataset are shown in Table 1. The ADC maps were calculated from DWI images with varying diffusion-weighting magnitude (b-value) (Equation \ref{eq:dwi}), using a mono-exponential fit for two b-values (b0, b1000) (Equation \ref{eq:dwi_log}) and least-square exponential fit for multiple b-values (b200, b600, b1000) (Equation \ref{eq:dwi_fit}). The Institutional Review Board approved this study, and informed consent was waived (Chang Gung IRB 202000609B0C501).

\begin{table}[htbp]
  \centering
  \label{tab:imaging-parameters}
  \caption{Imaging parameters of the cohort used in this study (n=207). The values are presented as the range from minimum to maximum values, or variables using different protocols within the dataset.}
  \begin{tabular}{p{5.9cm}p{4.6cm}p{4.6cm}}
    \toprule
    \textbf{Parameter} & \textbf{T\textsubscript{2}W MRI} & \textbf{DWI} \\
    \hline
    Sequence & Turbo Spin Echo (TSE) & Echo-Planar Imaging (EPI) \\
    Slice Orientation & Axial & Axial \\
    Acquired Matrix Size (read) & 224-320 & 128-172 \\
    Reconstructed Matrix Size (read) & 256-320 & 240-248 \\
    Reconstructed Pixel Spacing (mm$^2$) & 0.5$\times$0.5-0.8$\times$0.8 & 1.2$\times$1.2-1.4$\times$1.4 \\
    Slice Thickness (mm) & 4.0-7.5 & 4.0-5.0 \\
    Flip Angle (°) & 90 & 90 \\
    Echo Time (ms) & 80-101 & 60-80 \\
    Repetition Time (ms) & 3600-8060 & 3300-10844 \\
    Phase Encoding Direction & Anterior Posterior or Left Right & Anterior Posterior \\
    Pixel Bandwidth (Hz/pixel) & 190-200 & 1940-2441 \\
    b-values (s/mm$^2$) & - & [0,1000] or [200,600,1000] \\
    \bottomrule
  \end{tabular}
\end{table}

\begin{equation} \label{eq:dwi}
    S_{b_i} = S_{b_0} \cdot e^{(-b_i D)}
\end{equation}

\begin{equation} \label{eq:dwi_log}
    D = -\frac{1}{b_i} \ln \left(\frac{S_{b_i}}{S_{b_0}}\right)
\end{equation}

\begin{equation} \label{eq:dwi_fit}
D = -\frac{{N\sum_{i=1}^{N} b_i\ln(S_{b_i}) - \sum_{i=1}^{N} b_i \sum_{i=1}^{N} ln(S_{b_i})}}{{N\sum_{i=1}^{N} b_i^2 - (\sum_{i=1}^{N} b_i)^2}}
\end{equation}

where $b_i$ is the b-value, $S_{b_0}$ represents the signal intensity with no diffusion-weighting ($b=0$), $S_{b_i}$ is the signal intensity at $b_i$, $N$ is the number of b-values and $D$ denotes the ADC value.

\vspace{6mm} %6mm vertical space
\subsection{Network Architectures and Architectural Experiments}
In this study, we employed a residual U-Net model, a modified version of conventional U-Net \cite{ronneberger2015u}, as the benchmark for our segmentation framework. The model architecture was based on an encoder-decoder structure with symmetrical skip connections at each level. Each encoding level consisted of two residual blocks with 3D convolutional layers followed by instance normalization and parametric ReLU activation layer (PReLU). Downsampling was performed in the first convolutional layer (stride=2), and each upsampling block included two residual blocks with 3D strided transposed convolutions and skip connection concatenation layers. The model had four downsampling steps with kernel filters of 32, 64, 128, and 256 and a feature map depth of 512 in the bottleneck. The model was trained on (i) T\textsubscript{2}W MRI-only, (ii) T\textsubscript{2}W and ADC, and (iii) T\textsubscript{2}W, ADC, and b1000 DWI training data, with input channels set to 1, 2, and 3, respectively. Additional details on the model architecture can be found in Figure 1.

\begin{figure}[htbp!] 
\centering    
\includegraphics[width=0.8\textwidth]{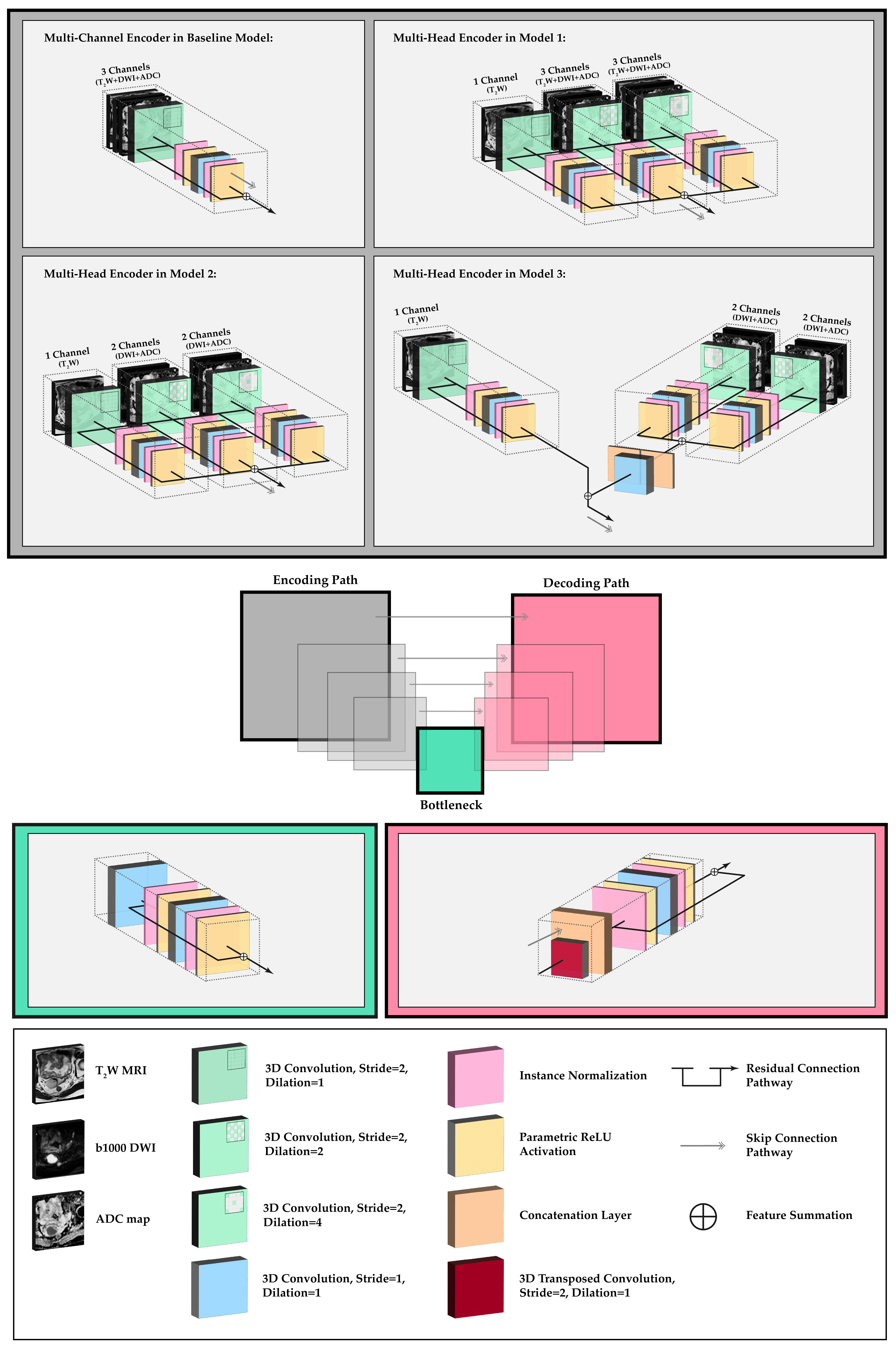}
\caption{The network topology and the multi-head segmentation training experiments. The operational blocks incorporate residual connections to facilitate the flow of information between different layers. The multi-head models use various encoding and weight-sharing configurations for T\textsubscript{2}W MRI and DWI using multiple heads with dilated convolution and connective residual operations.}
\label{fig:framework}
\end{figure}

To address the boundary effects of DWI in multi-channel training, we conducted a series of architectural experiments. In the first experiment, we replaced the first encoder block in the baseline multi-channel model with three encoding heads. The first head utilized a non-dilated 3D convolutional operation on T\textsubscript{2}W MRI inputs only. The other two heads included 3D convolutions with dilations of 2 and 4 respectively, accepting three input channels: T\textsubscript{2}W , b1000 DWI and ADC. These heads were connected by residual weight sharing and feature activation summation nodes, see Figure 1 (multi-head model 1).

In the second multi-head model, we utilized separate encoding of T\textsubscript{2}W MRI and DWI images while still maintaining weight sharing. The dilated heads in this model only included the DWI and ADC channels. This was performed to investigate the impact of DWI input channels on the overall contextual learning, see Figure 1 (multi-head model 2).

Finally, in the third experiment, we employed a multi-head encoding strategy with no weight sharing between the T\textsubscript{2}W and DWI heads. Similar to the second model, we applied dilated convolutions exclusively to the DWI images to facilitate independent contextual learning and reduce boundary attention. However, this architecture included an additional concatenation layer followed by a convolutional layer for dimensionality reduction and feature summation with the T\textsubscript{2}W head, see Figure 1 (multi-head model 3).

\vspace{2mm} %2mm vertical space

\subsection{Image Pre-processing and Implementation Details}
Prior to training, all mpMRI images were resampled to an in-plane resolution of 0.6$\times$0.6 mm\textsuperscript{2}, which was the most common T\textsubscript{2}W MRI resolution in the dataset (~85\%), and slice thickness of 4mm. Bilinear interpolation was applied for resampling T\textsubscript{2}W and b1000 DWI images, as it allows for smooth intensity transitions, preserving the details within the images. On the other hand, ADC images and contours were resampled using the nearest neighbor interpolation method. This method was chosen for its ability to retain the original discrete values of the images, which is crucial for ADC maps due to their quantitative nature, and for contours as they represent categorical labels or boundaries in segmentation tasks. Each channel was independently normalized to a mean of zero and unit variance. Finally, the dataset was randomly split into 157, 25, and 25 patients for training, validation, and testing, respectively.

During training, sub-volumes of size 256$\times$256$\times$16 were extracted stochastically from the training data, ensuring that each patch contained at least one annotated tumor slice. Random data augmentation operations, such as intensity shifting, scaling, and cropping, were applied to improve network generalizability. The models were trained for 100,000 iterations using the Dice loss function, which outperformed the combined Dice and cross-entropy and Tversky \cite{tversky1977features} losses during the initial experiments on input channels. The Adam optimizer with an initial learning rate of 1e-4 and weight decay of 1e-5 was used, and a cosine annealing learning rate scheduler was employed after each epoch. Validation was performed after each epoch, based on the Dice scores of whole image volumes, and the best performing weights were saved. Volumetric segmentations were generated using a sliding window algorithm with a 75\% overlap between adjacent patches. The models were evaluated using the Dice similarity coefficient (DSC), 95\textsuperscript{th} percentile Hausdorff distance (HD), mean surface distance (MSD), and percentage relative volume similarity metrics \cite{taha2015metrics}. PyTorch and Monai \cite{cardoso2022monai} DL libraries were used for all implementations.

\vspace{2mm} %2mm vertical space

\subsection{Channel Sensitivity Analysis and Visualization}
To assess the significance of individual channels in our models, we conducted a channel sensitivity analysis using sequential channel dropout. By setting each channel to zero one at a time, we compared the segmentation results obtained from the baseline multi-channel and proposed multi-head models to those achieved with no channel dropout. We employed the same quantitative metrics utilized in our previous analyses to perform a comprehensive comparison. To gain a more nuanced understanding of the importance of channels in our models, we employed a 3D version of Gradient-weighted Class Activation Mapping (GRAD-CAM) \cite{selvaraju2017grad}. This technique was utilized to generate saliency maps in the penultimate layers of our models, which highlighted the most relevant regions for segmentation. The application of this technique on the center-cropped patches extracted from the test image volumes provided a detailed and insightful visualization of the channel-wise feature importance.

\vspace{5mm} %5mm vertical space
%% ---- Results -----
\section{Results}
The baseline multi-channel model was evaluated on various input channels and loss functions. Multi-channel input trained with Dice loss achieved the best mean segmentation performance in terms of DSC across all test cases, with consistent results across input channels. The average DSC values for multi-channel (T\textsubscript{2}W, b1000 DWI, ADC) using Dice, combined Dice and cross-entropy, and Tversky losses were 0.672, 0.661, and 0.664, respectively (Figure 2).

\begin{figure}[htbp!] 
\centering    
\includegraphics[width=0.7\textwidth]{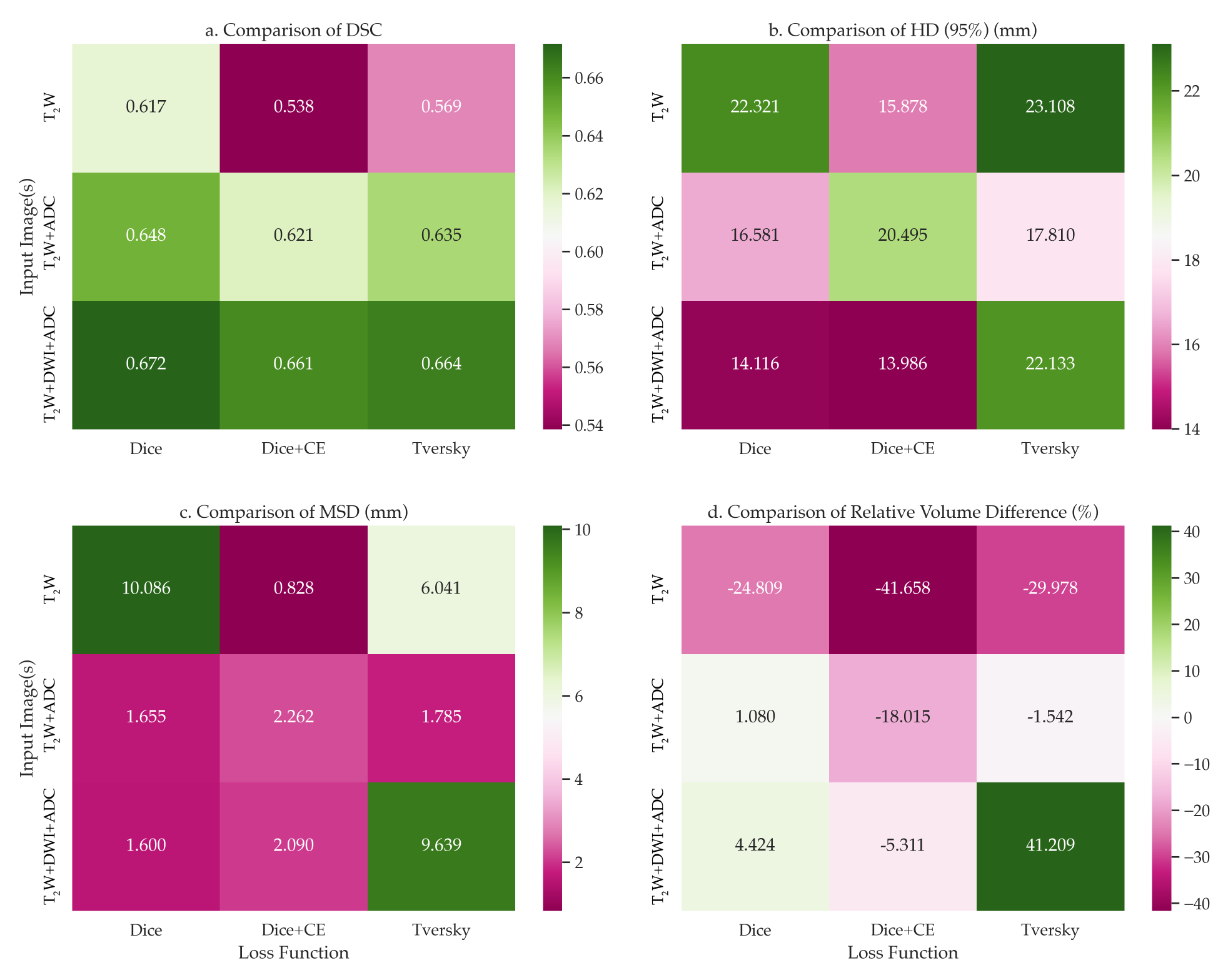}
\caption{Heatmaps of mean a. Dice similarity coefficient (DSC), b. 95\textsuperscript{th} percentile Hausdorff distance (HD), c. mean surface distance (MSD), and d. relative percentage volume difference for multi-channel residual U-Net model trained across different input images and loss functions from the test data. The model trained on all T\textsubscript{2}W, b1000 DWI, and ADC channels, and with Dice loss achieved the best overall segmentation performance across all experiments.}
\end{figure}

The comparison of segmentations obtained from the baseline multi-channel and the proposed multi-head models revealed that multi-head model 3 exhibited superior performance across the quantitative metrics analyzed (Figure 3a-d). Specifically, the median DSC values for the multi-channel model, multi-head model 1, multi-head model 2, and multi-head model 3 were 0.788 (confidence interval (CI), 0.568--0.776), 0.805 (CI, 0.538-0.769), 0.796 (CI, 0.537-0.776), and 0.823 (CI, 0.595-0.797), respectively (Figure 3a). However, the performance differences between the proposed multi-head models were not significantly different compared with the multi-channel model (p>0.05). On average, all models underestimated tumor volume compared with the contours drawn by the clinician, with the median relative percentage volume differences for each model of -14.4\%, -18.9\%, -9.7\%, and -12.0\%, respectively (Figure 3d). The multi-head model 3 demonstrated the best quantitative scores compared to the other experimental architectures, therefore, the segmentation contours predicted by this model were compared with those obtained from the baseline multi-channel model (Figure 4).

% \vspace{6mm} %6mm vertical space

\begin{figure}[htbp!] 
\centering    
\includegraphics[width=1.0\textwidth]{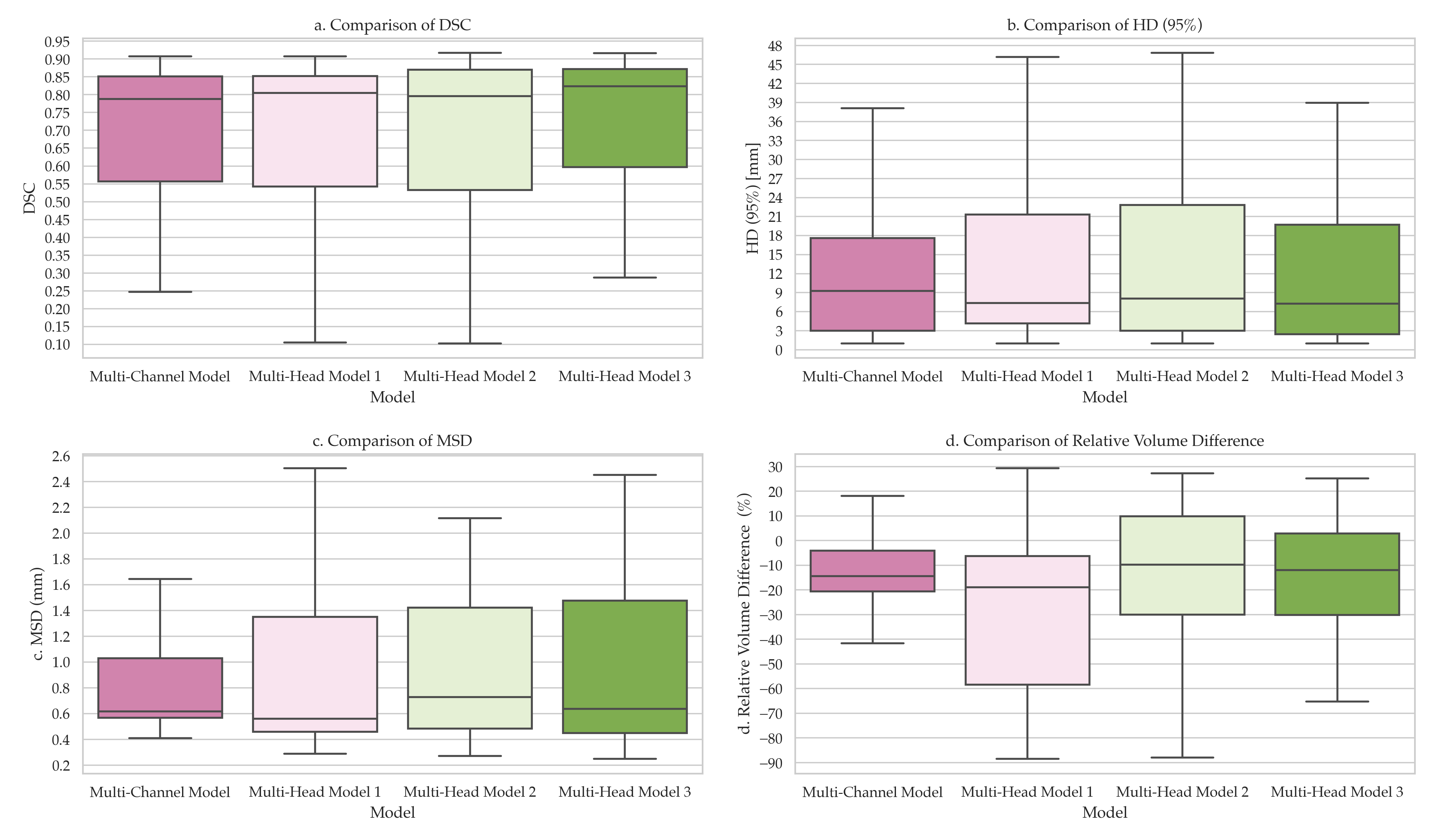}
\caption{Comparison of segmentation performance based on a. Dice similarity coefficient (DSC), b. 95\textsuperscript{th} percentile Hausdorff distance (HD), c. mean surface distance (MSD), and d. relative percentage volume difference between the baseline multi-channel model and three variations of the dilated multi-head model. Overall, the multi-head model with dilated convolutions and separate DWI/ADC feature aggregation (multi-head model 3) achieved the best performance across all models.}
\label{fig:models boxplots}
\end{figure}

% \vspace{8mm} %8mm vertical space

\begin{figure}[htbp!] 
\centering    
\includegraphics[width=0.75\textwidth]{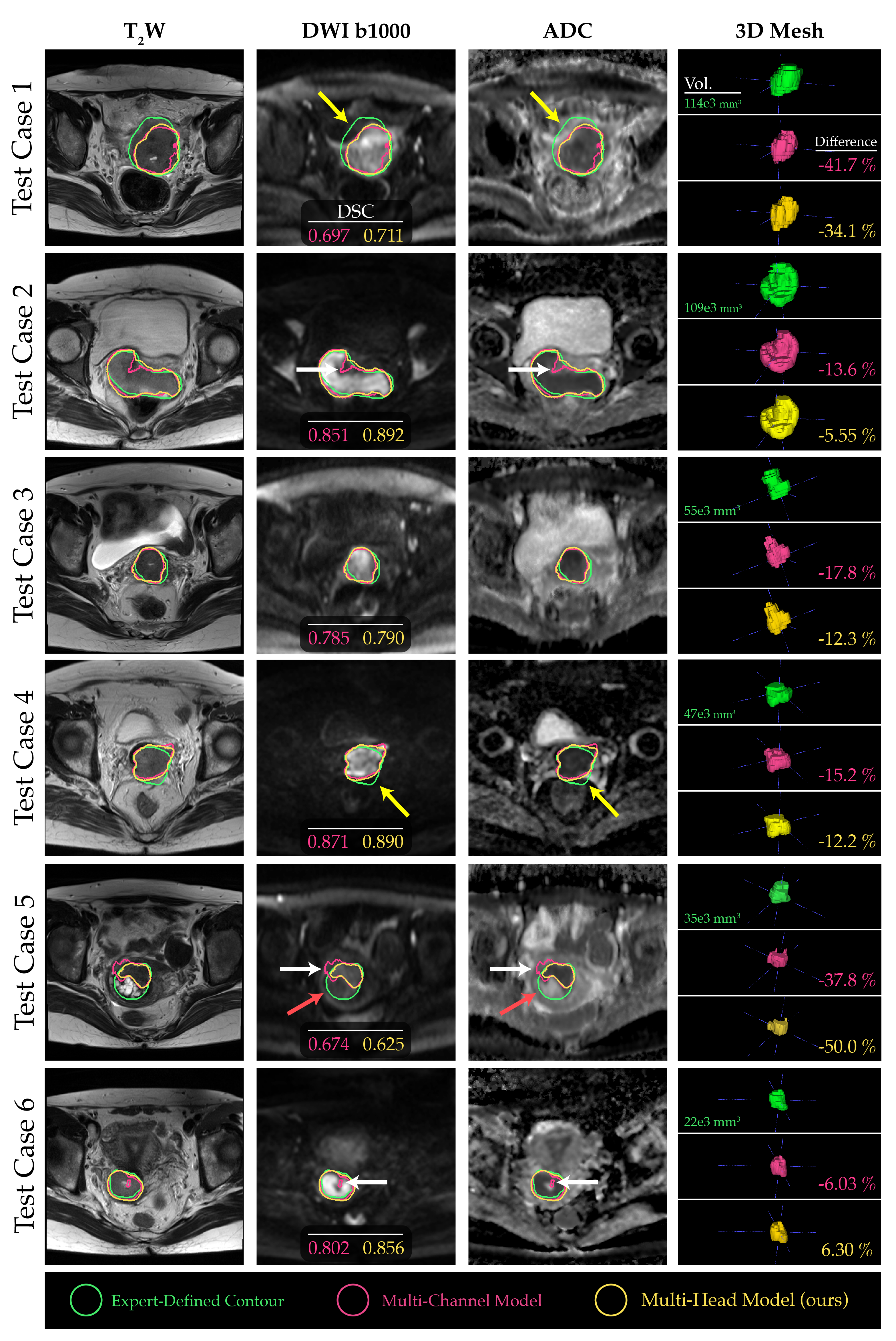}
\caption{Comparative evaluation of the baseline multi-channel and proposed multi-head models for 6 test cases, arranged in descending order of tumor size. The yellow arrows indicate regions where distortion in DWI and subjectivity in tumor location make contour propagation from T\textsubscript{2}W images challenging. Conversely, the white arrows highlight regions where the proposed model performed better by capturing boundaries and contextual information compared to the baseline multi-channel model. The red arrows highlight tumor regions within the ground truth contours that do not exhibit strong impeded diffusion, thus remaining undetected by the DL models.}
\label{fig:seg results}
\end{figure}

The channel sensitivity analysis conducted between the multi-channel and proposed multi-head model indicated that the multi-head model's tumor segmentation performance significantly suffered when the ADC channel was removed (Figure 5). Moreover, the saliency maps generated from both models revealed that b1000 DWI images had a relatively minor impact on the overall tumor segmentation performance for larger tumor volumes, while T\textsubscript{2}W and ADC images were more crucial (see Figure 6, test cases 1 and 2). Conversely, for smaller and more challenging tumor masses, the absence of ADC values had a more pronounced adverse effect on the final outcome (see Figure 6, test cases 3-5).

\begin{figure}[htbp!] 
\centering    
\includegraphics[width=1.0\textwidth]{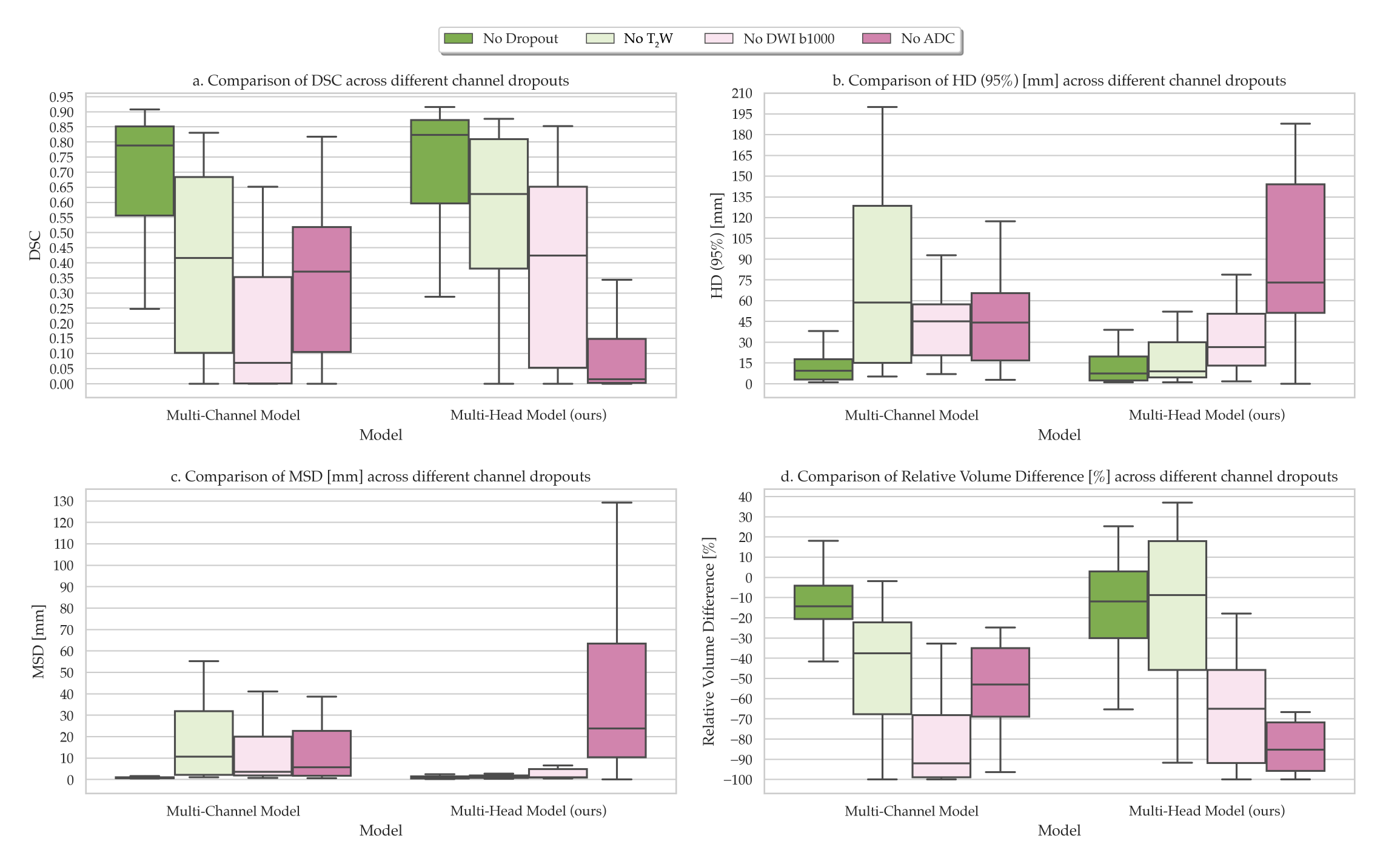}
\caption{Comparison of the sensitivity of the multi-channel and proposed multi-head models to different input channels based on quantitative metrics, including a. Dice similarity coefficient (DSC), b. 95\textsuperscript{th} percentile Hausdorff distance (HD), c. mean surface distance (MSD), and d. relative percentage volume difference. Our channel dropout analysis revealed a strong dependence of the proposed multi-head model on the ADC channel, indicating its importance in achieving accurate segmentation performance.}
\label{fig:sensitivity_boxplots}
\end{figure}

\begin{figure}[htbp!] 
\centering    
\includegraphics[width=0.72\textwidth]{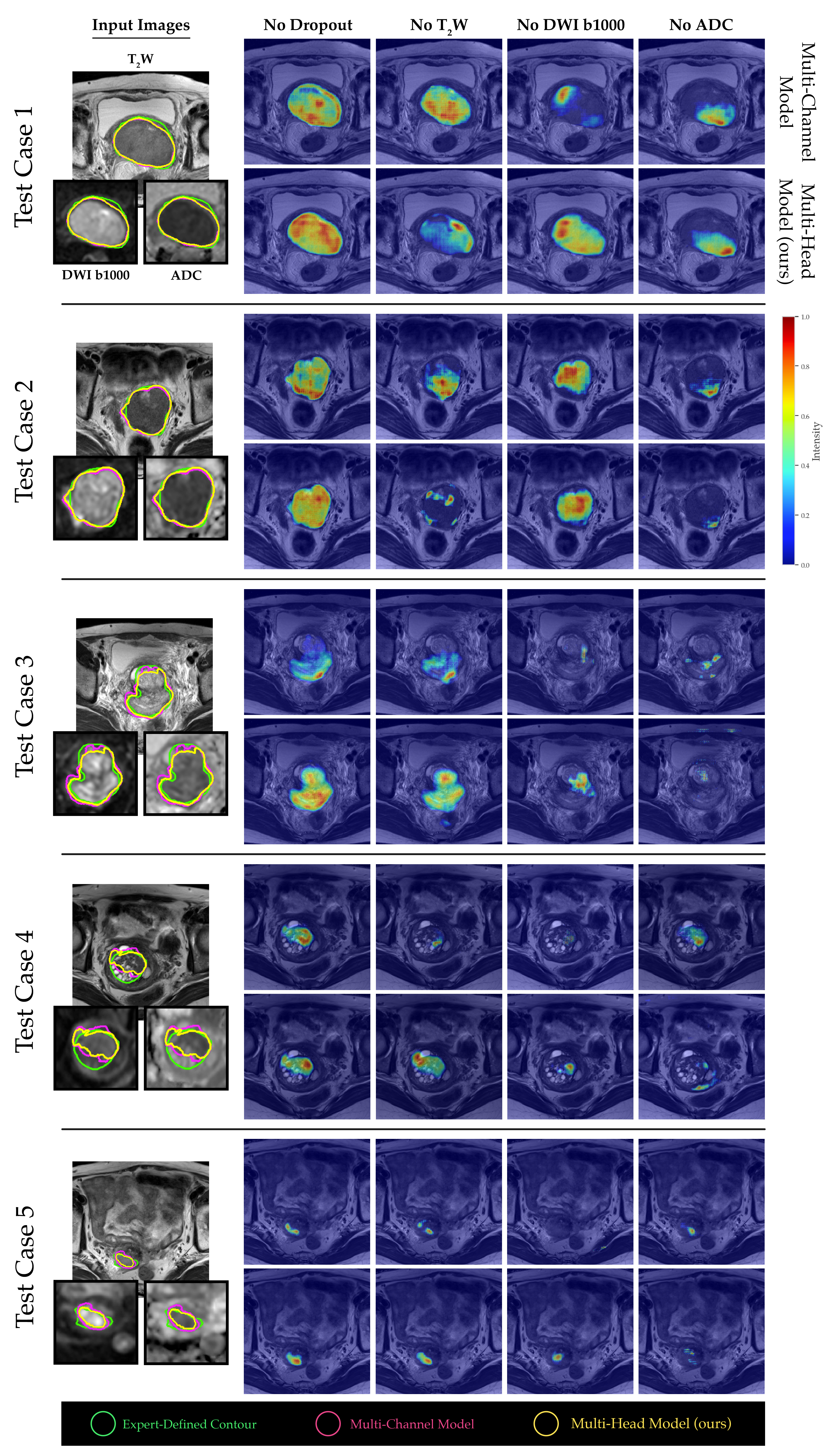}
\caption{The normalized 3D GRAD-CAM saliency maps generated from the penultimate layer of models, highlighting the most relevant regions for segmentation under different channel dropout conditions at test time. To improve visualization, the b1000 DWI and ADC images for smaller tumors were scaled.}
\label{fig:activations}
\end{figure}

% \FloatBarrier
\section{Discussion}
DWI is a critical functional imaging technique for the detection and localization of tumors. This is particularly evident on ADC maps, where regions depicting impeded water molecule diffusion are indicative of increased cell density, often signifying a more aggressive disease \cite{mcveigh2008diffusion}. However, using DWI and ADC maps in conjunction with T\textsubscript{2}W MRI  sequences presents several challenges, such as voxel misalignment due to distortion and soft-tissue deformations between scans, variations in tumor delineations between the two modalities and the absence of standardized protocols for mpMRI. In this study, we proposed a multi-head framework that utilizes both DWI (including ADC) and T\textsubscript{2}W MRI for cervical cancer segmentation, a process well-suite for biomarker extraction. Our framework includes separate encoding heads that extract contextual information about tumors using dilated (or atrous) convolutions and shared residual connections. We have demonstrated that our technique provides more robust and boundary-aware segmentation of cervical tumors when compared with the baseline multi-channel architecture used in previous studies \cite{lin2020deep, hodneland2022fully}. Our findings using the multi-channel training approach are comparable to previous reports on segmentation of cervical tumors on multiparametric MRI \cite{wang2021multimodal, lin2020deep, hodneland2022fully}. However, it is challenging to make direct comparisons of our results due to the lack of public databases for cervical cancer segmentation on MRI \cite{kalantar2021automatic} and the use of different datasets.

Dilated convolutions are a crucial component of several successful segmentation architectures, including DeepLabv3+ \cite{chen2018encoder}, DeepLab \cite{chen2017deeplab}, residual enhanced U-Net and atrous convolution (AtResU-Net) \cite{shah2021colorectal}, and 3D deeply supervised fully-convolutional network with concatenated atrous convolution (3D DSA-FCN) \cite{wang2019automatic}. While most of these techniques utilize dilated convolutions throughout their architecture's encoding steps, we propose a multi-head framework that uses these operations for separate contextual and representational learning in DWI and ADC images only for the first block of the encoder. This approach ensures that the training parameters are not substantially increased compared with baseline multi-channel architectures, and the model is easier to interpret with conventional methods for future explainability studies. Moreover, lighter networks are better suited for online MR-guided treatments, where segmentation and planning are performed live on the scan of the day before radiation treatment \cite{breto2022deep}, and speed is of the essence. In this study, we trained our models using anisotropic sub-volumes to maintain a greater focus on the plane of acquisition (axial) for 2D MRI. However, our method can be extended to 3D MRI scans, which are more commonly used for radiation therapy.

Channel sensitivity and saliency mapping of our experimental model indicated that our algorithm was more sensitive to DWI and, particularly, ADC maps, which potentially make it more robust to changes in acquisition protocol on MRI scanners. This approach could also serve as a strategy for more generalizable and cross-disease detection models \cite{lin2023generalizable}. However, this dependence on DWI, as demonstrated in this study, may result in underestimation of the predicted tumor volume for malignancies with heterogeneous tumor mass diffusion. Moreover, the subjectivity associated with inter-operator variability presents another drawback, with more reliable segmentations only attainable through the use of consensus contours. The decision to include specific areas within the ground truth contour of the tumor is a discretionary choice made by the clinical expert annotating the images, a decision that relies heavily on their professional training and experience. Hence, future studies should aim to employ consensus ground truth contours and evaluate the segmentation outcome through a number of expert human reader assessments to ensure the accuracy and reliability of the results for clinical decision-making.

In conclusion, our proposed multi-head framework that combines DWI, ADC and T\textsubscript{2}W MRI for cervical cancer segmentation has demonstrated improved accuracy and robustness compared to conventional multi-channel architectures. The use of dilated convolutions in only the first block of the encoder improves contextual learning with no significant parameter increase compared with conventional U-Net models. However, the dependence on DWI and subjectivity in inter-operator variability are potential limitations that need to be addressed in future studies through the use of consensus ground truth contours and expert human reader assessments. Overall, our approach has the potential to improve clinical decision-making for not only cervical cancer but also other pelvic malignancies.

\section*{Acknowledgments}
This study represents independent research funded by the National Institute for Health and Care Research (NIHR) Biomedical Research Centre and the Clinical Research Facility in Imaging at The Royal Marsden NHS Foundation Trust and The Institute of Cancer Research, London. The views expressed are those of the author(s) and not necessarily those of the NIHR or the Department of Health and Social Care. Gigin Lin received research funding from the Ministry of Science and Technology Taiwan (MOST 110-2628-B-182A-018).

%Bibliography
\bibliographystyle{unsrt}  
\bibliography{references}  

\end{document}